\documentclass[conference]{IEEEtran}
\usepackage{amssymb,amsmath,amsthm}
\usepackage{caption}
\usepackage{mdwmath}
\usepackage{mdwtab}
\usepackage{soul}
\usepackage{cite}
\usepackage{multirow}
\usepackage{color} 
\usepackage{setspace}
\usepackage{esvect}
\usepackage{amsmath}

\usepackage{subfig}

\ifCLASSINFOpdf
  \usepackage[pdftex]{graphicx}
  \graphicspath{{./}}
  \DeclareGraphicsExtensions{.pdf,.jpeg,.png,.xps}
\else
\fi

\begin{document}

\title{DiffFlow: Differentiating Short and Long Flows for Load Balancing in Data Center Networks}

\author{\IEEEauthorblockN{Francisco Carpio, Anna Engelmann and Admela Jukan}
	\IEEEauthorblockA{Technische Universit{\"a}t Braunschweig, Germany}
	\IEEEauthorblockA{Email:\{f.carpio, a.engelmann, a.jukan\}@tu-bs.de}
}

\maketitle

\begin{abstract}
In current Data Center Networks (DCNs), Equal-Cost MultiPath (ECMP) is used as the de-facto routing protocol. However, ECMP does not differentiate between short and long flows, the two main categories of flows depending on their duration (lifetime). This issue causes hot-spots in the network, affecting negatively the Flow Completion Time (FCT) and the throughput, the two key performance metrics in data center networks.  Previous work on load balancing proposed solutions such as splitting long flows into short flows, using per-packet forwarding approaches, and isolating the paths of short and long flows. We propose DiffFlow, a new load balancing solution which detects long flows and forwards packets using Random Packet Spraying (RPS) with help of SDN, whereas the flows with small duration are forwarded with ECMP by default. The use of ECMP for short flows is reasonable, as it does not create the out-of-order problem; at the same time, RPS for long flows can efficiently help to load balancing the entire network, given that long flows represent most of the traffic in DCNs. The results show that our DiffFlow solution outperforms both the individual usage of either RPS or ECMP, while the overall throughput achieved is maintained at the level comparable to RPS.
\end{abstract}


% \fi
\IEEEpeerreviewmaketitle

\section{Introduction}

%% Scenario
\par In recent years, the number of Internet applications hosted in Data Centers has been fast growing, making the Data Center Network (DCN) operation ever more complex. Generally, two different categories of traffic can be found in DCN: (1) those associated with user tasks, e.g., web browsing or search queries, and, (2) those generated by virtual machine migration, data backup, or MapReduce operations. The first category, referred to as short flows, includes flows generated by user tasks that have short duration and need to be transmitted before the so-called \emph{Flow Completion Time (FCT)}, specified in the Service Level Agreement (SLA) with users.  The second category of flows, referred to as long flows, are generated by applications with long duration and typically require an adequate \emph{throughput}, i.e.,  higher than the minimum acceptable, however without stringent temporal constraints.

%% Problem 
\par
\begin{figure*}[!t]
	\centering
	\subfloat[ECMP bottleneck problem]{\includegraphics[width=2.5in]{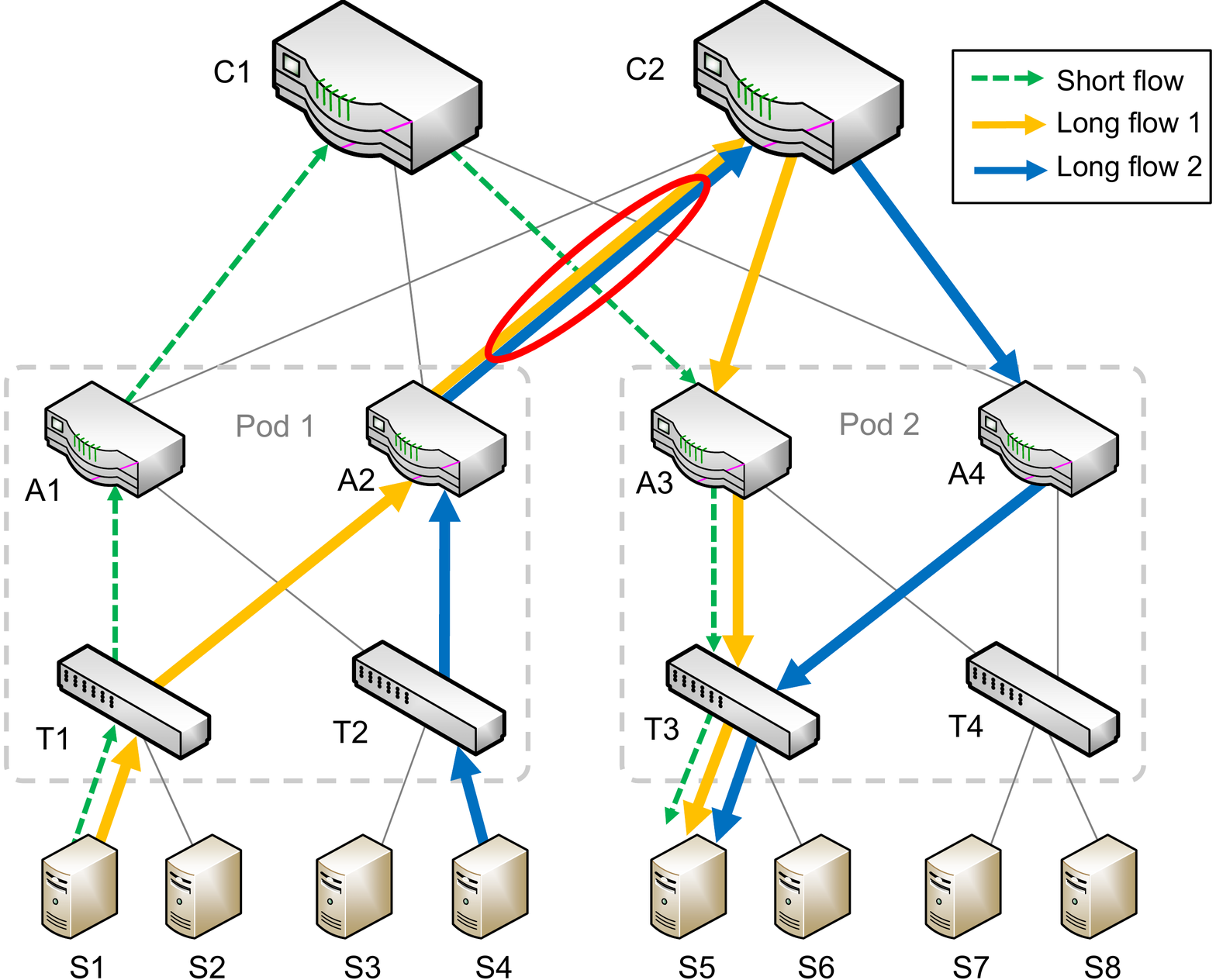}
		\label{ecmp}}
	\hfil
	\subfloat[DiffFlow operation]{\includegraphics[width=2.5in]{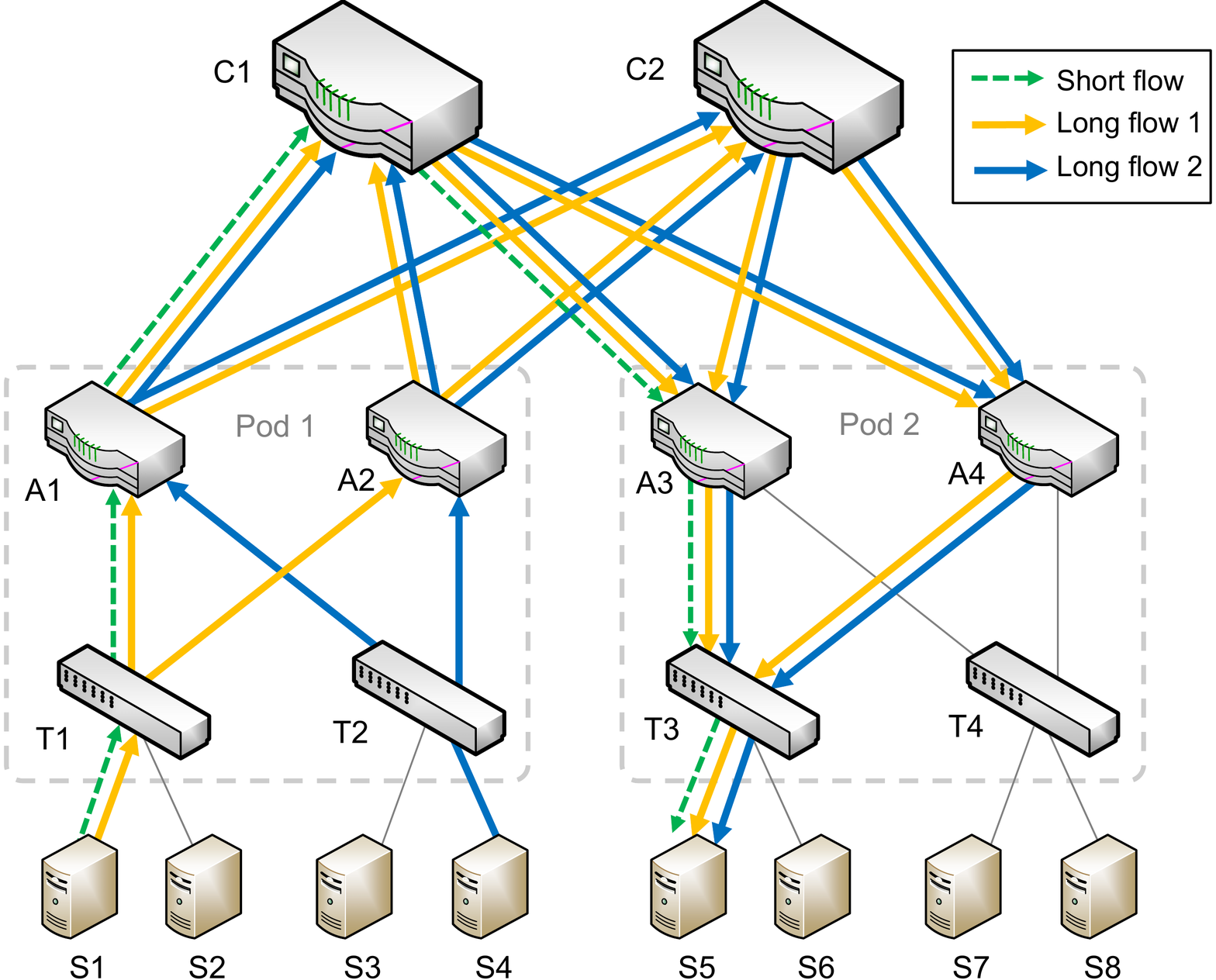}
		\label{hybrid}}
	\caption{Comparison of ECMP and DiffFlow}
	\label{operation}
\vspace{-0.4cm}
\end{figure*}

 Mixing these two types of flows over the same network with their contradictory requirements for fast FCT and high throughput has already been identified as a challenge. Traditionally, Equal Cost Multi Path (ECMP) is used as default routing algorithm in data centers, where an individual flow between a pair of servers is routed over one possible shortest path, which is selected calculating the hash value of the 5-tuple header fields. With ECMP, two long flows can end up being routed over the same path causing hot-spots in the network, as illustrated in Fig. \ref{ecmp}. As a consequence, the throughput decreases and the path latency traversing the congested link increases, increasing the FCT. This in turn affects the user experience when short flows associated to user tasks are routed over congested paths. Previous work has proposed three general categories of solutions: (1) sacrificing one of the two constraints to optimize the other, (2) increasing the complexity of traffic engineering with priority scheduling, and (3) modifying the current DCN architectures, e.g., modifications of TCP stack, using OpenFlow switches. 

%% solution 
\par In this paper, we present a new, hybrid category of solutions for mixed short and long flows in data center networks called DiffFlow, capable of achieving the desired trade-off between low latency (fast FCT) and high throughput. In our approach, we propose to use traditional ECMP for short flows, minimizing the FCT with a negligible out-of-order problem typically present in per-packet basis approaches, while for long flows, we propose to use Random Packet Switching (RPS), capable of load balancing without affecting the path latencies of short flows, as shown in Fig. \ref{hybrid}. Since long flows typically constitute  80-90\% of total DCN traffic, while 80\% of all flows are short, and smaller than 10KB in size \cite{Benson2010}, this means, that with our method, 80\% of flows are routed with ECMP by default, while the remaining 20\% of flows with a long duration use RPS. For our idea to work, we propose the use of packet sampling technique on OpenFlow switches to detect long flows and the SDN controller to advertise about the presence of long flows and apply RPS rules on the switches. We show analytically and with simulations, that the proposed method can effectively load balance the network, while keeping FCT and throughput within the pre-defined ranges.

\par The remainder of this paper is organized as follows. Section II presents related work. Section III presents the design of our hybrid method. In Section IV, the theoretical analysis is presented. In Section V, we discuss the performance evaluation, and finally in Section VI we present the conclusions.

\section{Related work}
\par The general classification of different requirements and solutions for short and long flows in DCN has been discussed in \cite{Rojas-Cessa2015}. Here, we provide an overview of solutions related specifically to multipath forwarding schemes in DCNs.

\par One of the first  solutions proposed to address the long flow problem was Hedera \cite{Al-Fares2010}. Proposed was the use of a centralized flow scheduling, using OpenFlow switches, to relocate long flows for load balancing. Because of the use of central controller, the algorithm takes some time for the reallocation of flows which was shown to make Hedera slow reacting dynamically to changes in traffic patterns.  

\par MPTCP, an extension of TCP for DCN, has been analyzed in \cite{Raiciu2011}. In this scheme, the end hosts are responsible to split flows into sub-flows and send them over different paths. However, the modification of the TCP/IP stack is a practical challenge. To avoid modification of TCP, the authors of CONGA \cite{Alizadeh2014} proposed a distributed, congestion-aware and load balancing scheme for DCNs. CONGA splits flows in flowlets and depending on the estimation of path congestions based on feedbacks from the switches, choose the optimum path. For this scheme to work, however, the custom switching ASICs are required to maintain the congestion tables in the switches. Another approach that does not require TCP modification is  RepFlow \cite{Xu2014a}, which uses replication of short flows. This scheme uses two independent TCP connections differentiated by different port numbers, whereby the hashing performed by ECMP provides two different paths, one for original short flow and another for the replication. Since the amount of data carried by short flows is insignificant compared to long flows, the overhead produced by this solution is negligible. However, since this approach is not a congestion control scheme, the collision of long flows can still occur, which was addressed by FreeWay \cite{Wang2016}. Here, the paths are differentiated between the low latency paths and high throughput paths. Short flows are transmitted over low latency paths using ECMP and FreeWay allocates long flows over different high throughput paths. In this way, they achieve isolation between short and long flows avoiding conflicts and optimizing latency and throughput. 

\par Our solution was motivated by TinyFlow \cite{Xu2014} and RPS \cite{Dixit2013, Kaymak2015}. The former (TinyFlow) splits long flows into short flows, and forwards them randomly by making use of ECMP. In order to detect long flows, OpenFlow switches perform sampling periodically. When two samples of the same flow are detected, the dynamic random re-routing algorithm changes the egress port of the switch. The latter method, i.e., RPS, uses random packet spraying technique to forward packets through multiple shortest paths. Unlike ECMP, RPS forwards each packet individually ("spraying") to the egress ports at the DCN switches. (Although this feature is not enabled by default, commodity switches can perform it.) The main drawback of this method is the out-of-order problem for TCP. As the authors prove, however, under the symmetry assumption in fat tree topologies and traffic patterns, this problem can be minimized, and even neglected. Also, custom queue management scheme can be applied to minimize the differences on path latencies. 

\begin{figure*}
	\centering
	\includegraphics[width=\textwidth]{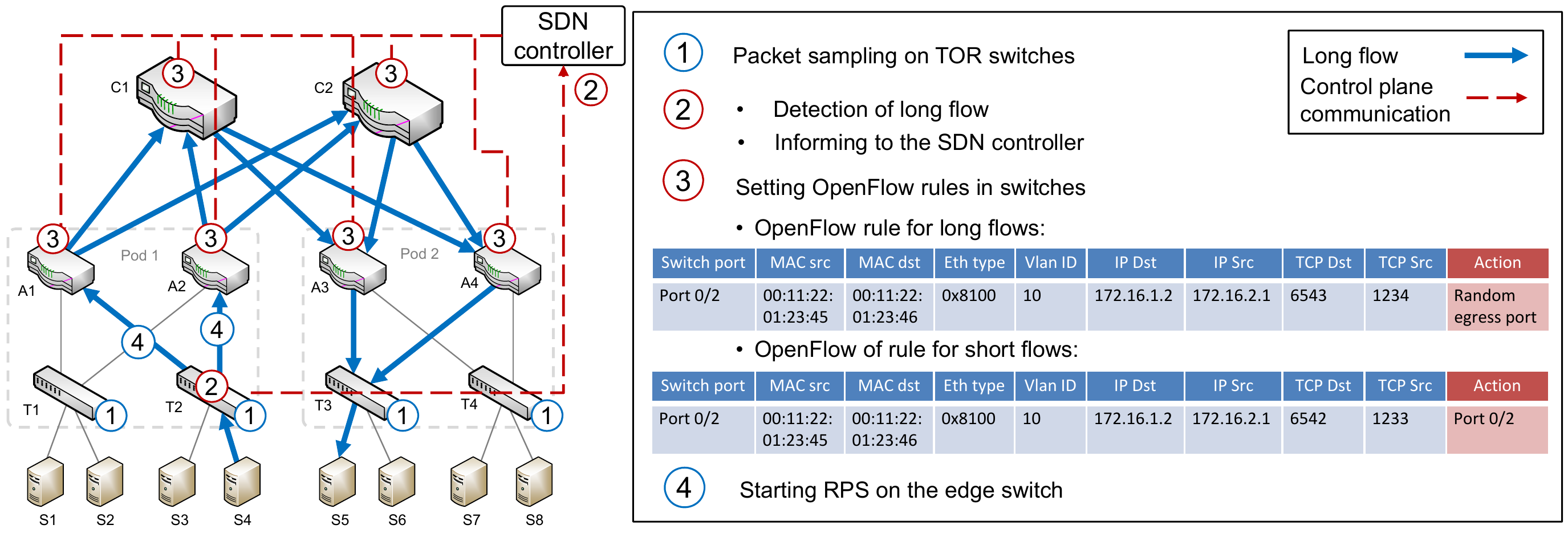}
	\caption{DiffFlow operation}
	\label{procedure}
\vspace{-0.4cm}
\end{figure*}

\par In spite of the ideas presented in these works, current data centers are slow to adopting new schemes, especially when they require modification of legacy network elements, or transport protocols. For that reason, our focus is on the solution deploying SDN, which is well-adopted in data centers. In our proposal, OpenFlow switches are essential to differentiate the flows. In our model, we use the packet sampling technique, also widely adopted in practice, such as sFlow \cite{sFlow}. The main advantage of this method compared to others is that the sampling of packets is only realized locally in the TORs OpenFlow switches. Therefore, the SDN controller is not responsible for this task, assuring a high scalability for bigger networks, which is addressing an important concern of centralized SDN controllers. Packet sampling is however not the only detection technique that can be used, and our solution is open to other methods, for instance, per-flow statistics as in Hedera \cite{Al-Fares2010}, or end-host based monitoring as in Mahout \cite{Curtis2011}. On the other hand, it is important to underline that DiffFlow can coexist with traditional TCP protocol or newer versions, since its procedure is transparent to upper layer protocols.

% It should be noted that our solution works in a distributed manner, and the SDN controller is used only to discover the presence of long flows. 

\section{System Design}

In this Section, we show the operation of DiffFlow, and define the performance metrics used in our solution.

\subsection{Operation}

\par Fig. \ref{procedure} shows a typical DCN type of topology, where at lowest level there are 8 servers (S1-S8) connected in pairs to the TORs switches (T1-T4), forming two independent pods jointly with the aggregation switches (A1-A4).  At the highest level, all aggregation switches connect to all core switches (C1, C2). 

\par Let us now illustrate the DiffFlow operation in four basic steps. We start with network system where there are only short flows at first. The switches forward of flows by the default routing protocol, i.e., ECMP, and without the intervention of the SDN controller. At the same time, however, TORs switches are sampling the packets every certain pre-defined period of time, in order to detect when a long flow enters to the network. Based on the length of the sampling period, the flow duration can be labeled as short or long. For instance, if we set a sampling period with a small value, flows with shorter duration will be marked as long flows. Let us now assume that after a series of short flows, long flow is to be detected, between the servers S4 and S5. Here, the switch T2 starts forwarding the first packets of the flow as if they were from a short flow, while the packet sampling process is running in the background (Step 1). When the packet sampling process detects two packets belonging to the same flow, indicating a potential congestion, T2 informs the SDN controller about the detection of a long flow, sending the MAC and IP addresses, as well as the TCP ports that are read from the packets (Step 2). The SDN controller saves this information and advertises it to all aggregation and core switches (A1-A4, C1, C2) sending the corresponding OpenFlow rules (Step 3). 

\par Let us look into more detail how OpenFlow sets the rules on the example of switch A2. The first rule (long flows) specifies that all the packets received at \textit{port 0/2}, with source MAC \textit{00:11:22:01:23:45} belonging to one interface of S4, destination MAC \textit{00:11:22:01:23:46} belonging to one interface of S5, source IP address \textit{172.16.1.2} associated to server S4, destination IP address \textit{172.16.2.1} associated to server S5, source TCP port \textit{6543} associated to the long flow connection of the sender S4 and destination TCP port \textit{1234} of the receiver S5, have to be forwarded towards a random egress port (Action field). On the other hand, other packets that do not match with this rule will follow other automatically generated rules (by hashing of 5-tuple header fields), such as in the next example of the rules (short flows). In the latter case (short flows), the rule is set up to forward packets in a short flow also between S4 and S5. The MAC and IP addresses are hence the same as previously,  but the different TCP ports specify that it is another TCP connection. Here, the action field specifies a concrete egress port, since this rule applies to a short flow and therefore all the packets have to follow the same path. Once the rules are established in all intermediate switches, the SDN signals to T2 to start applying RPS in the long flow (Step 4). From this point on, when an intermediate switch receives packets, specific rules (automatically generated, or specified by the controller) apply the required action to all the packets of the said TCP connection.

\par As it is well known that using optimal path computation in a centralized controller is not scalable in DCNs due to the high quantity of requests, it should be noted that our solution does not have this problem. This is because the SDN controller is only responsible for the advertisement of the long flows to the switches. Furthermore, the communication with the controller only takes place for long flows, which makes our solution scalable even for large topologies, since long flows are only about 10\% of all flows in DCN \cite{Benson2010} .

\begin{figure*}[!t]
	\centering
	\subfloat[ECMP]{\includegraphics[width=2.5in]{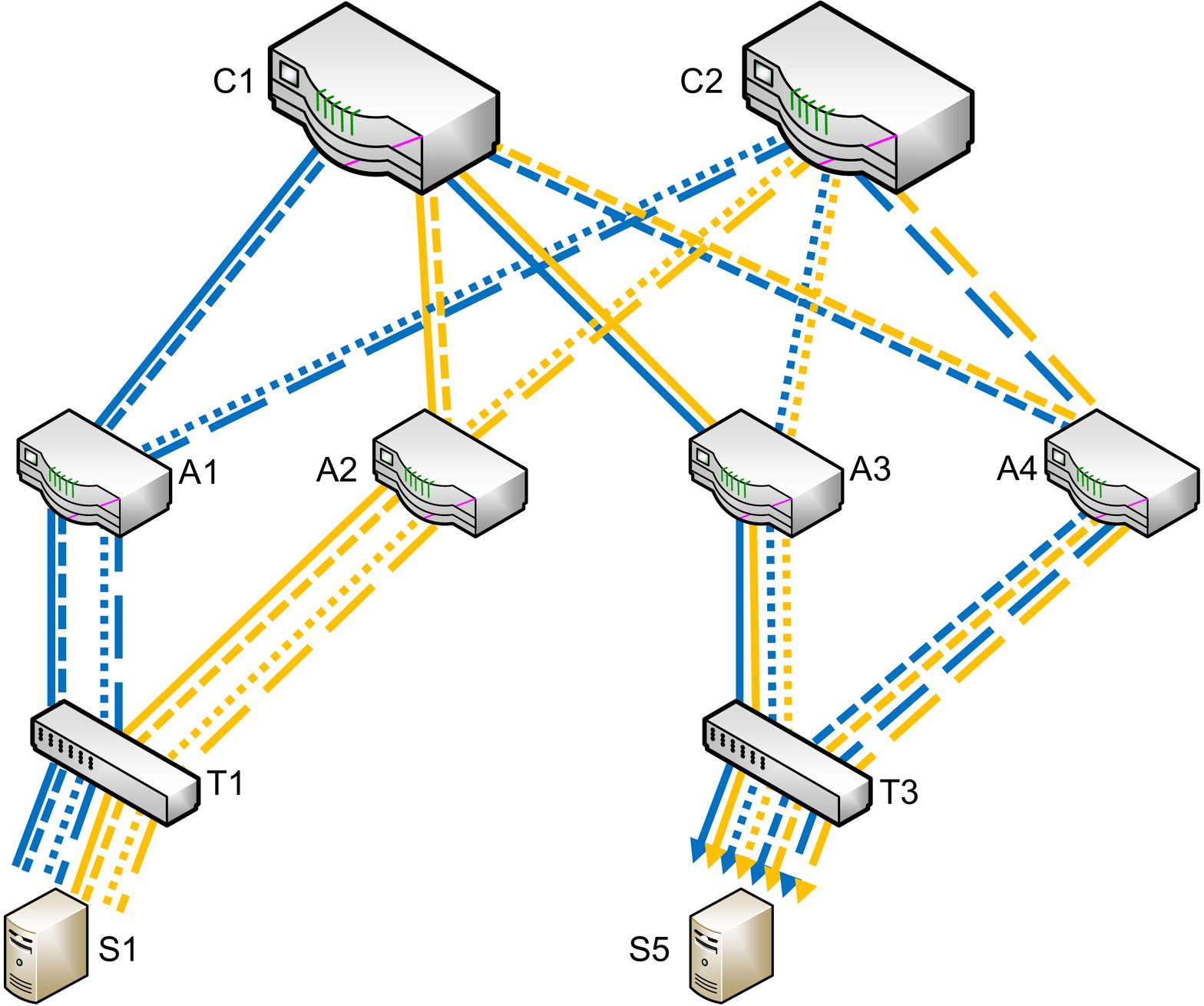}
		\label{ECMP-paths}}
	\hfil
	\subfloat[RPS]{\includegraphics[width=2.5in]{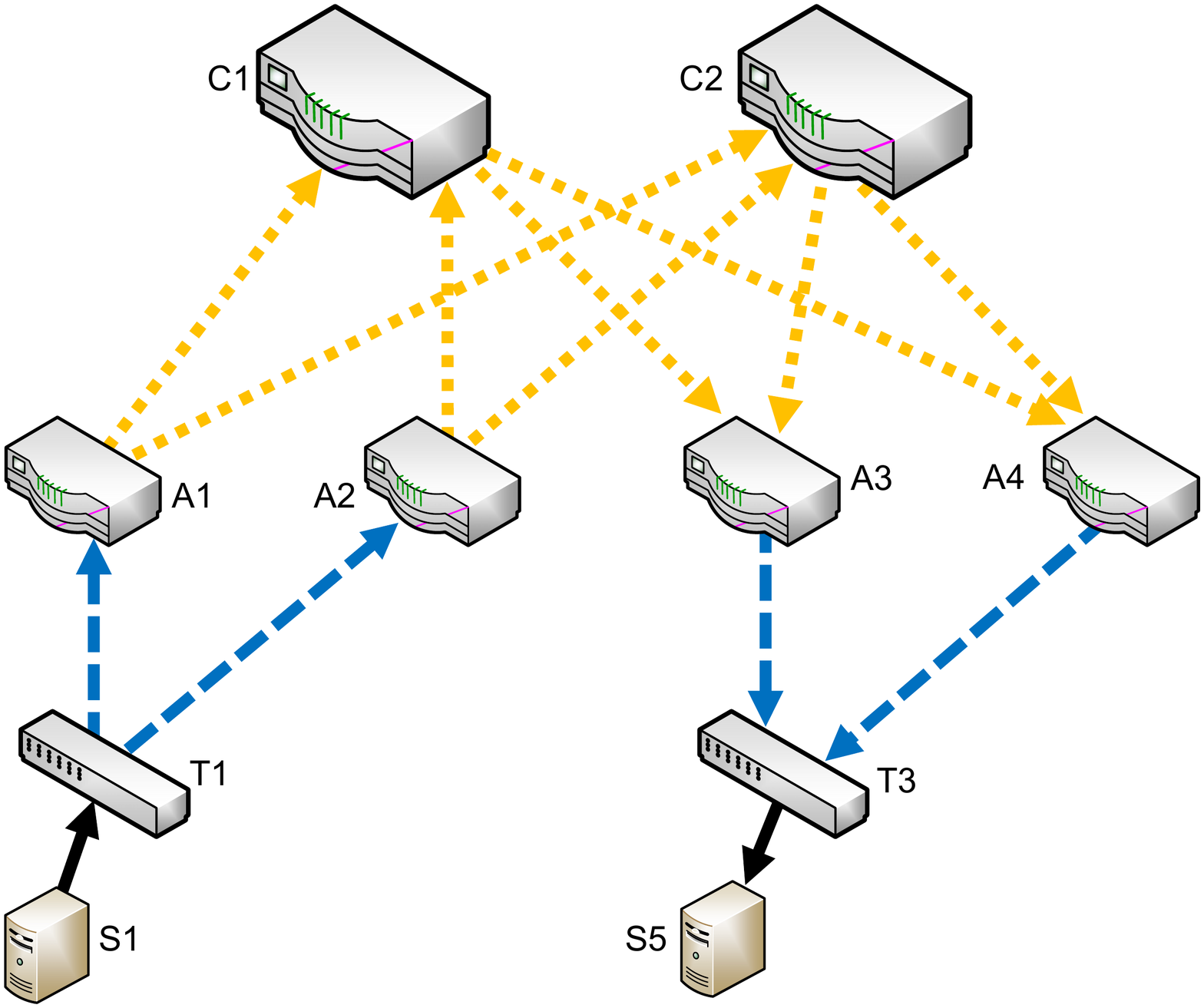}
		\label{RPS-links}}
	\caption{Forwarding from source to a destination at different pod}
	\label{paths}
\vspace{-0.4cm}
\end{figure*}

\subsection{Performance metrics} \label{performance_metrics}

\par The performance of our model is measured by two main metrics, the Flow Completion Time (FCT) and throughput. Both metrics are used for all flows, however the goal is to minimize FCT for short flows, while maintaining an acceptable throughput for long flows. 

\par The FCT of a flow is defined as the difference between two time stamps, the first time stamp when the first packet of a flow leaves a source server, and the second time stamp when the last packet of the same flow arrives at the destination server. If we assume that the network is empty (no queues at the switches), the FCT between two servers is the best achievable ("ideal"), where only the service times of the intermediate nodes and the transmission time of the packets are considered. We use the normalized value of this metric, defined as the ratio between the measured value when there are no ideal conditions, i.e., in case of queuing time affecting the FCT, and the ideal FCT.

\par The throughput of a flow is calculated by counting the percentage of packets dropped in the queues. In order to simplify this calculation, we model a path used by a flow as a M/D/1/K system with $\lambda$ packets per second and $\mu$ service time. Therefore, the ideal case is when there are no packets dropped due to blocking on intermediate switches, matching the load of a flow with the maximum achieved throughput. Knowing the load as an input parameter, we only need to count the number of dropped packets to calculate the blocking probability and compute the measured throughput. Also here, we normalize the the measured throughput with the ideal throughput.

\section{Analysis} \label{analysis}

In this Section, we analyze the mean FCT values when, given a certain flow, all packets follow the same randomly chosen path (i.e. ECMP) and when every packet is randomly sprayed over all possible paths (i.e. RPS). Then the results of our analysis, in Section \ref{performance}, will show the behavior of ECMP and RPS in comparison with our DiffFlow proposal, which combines both schemes. All the possible paths that a certain flow can choose in the ECMP case are shown in Fig. \ref{ECMP-paths}, while all the possible paths that a packet of a flow can choose in the RPS case are shown in Fig. \ref{RPS-links}. The notation for the analysis is provided in Table \ref{notation}.

\begin{table}[!t]
	\renewcommand{\arraystretch}{1.3}
	\caption{Notation}
	\label{notation}
	\vspace{-0.1cm}
	\centering
	\begin{tabular}{l l}
		\hline
		\textbf{Parameter} & \textbf{Meaning}\\
		\hline
%		$V$  & set of nodes\\
%		$E$  & set of links\\
%		$G (V, E)=G$ & acyclic graph\\
		$s,d$ & source and destination node\\
		$\mathcal E_{in} (v)$ & set of incoming links of node $v$\\
		$\mathcal E_{out} (v)$ & set of outgoing links of node $v$\\
%		$e_{vv'}$ & link from $v$ to $v'$\\
%		$c_{e_{vv'}}$ & link capacity from $v$ to $v'$\\
		$\vv{p}=\{p_0,p_1, ..., p_{N-1} \}$ & set of all paths between $s$ and $d$\\
		$\vv{P}=\{P_0,P_1, ..., P_{N-1} \}$  & set of path probabilities of $\vv{p}$\\
		$\vv{\tau}=\{\tau_0,\tau_1, ..., \tau_{N-1} \}$  & set of end-to-end delays of $\vv{p}$ \\
		$\mathfrak y_{l}$& number of fwd nodes on path $p_l$\\
		\hline
	\end{tabular}
\vspace{-0.5cm}
\end{table}

\par In our model, we assume that all links have the same capacity and all $N$ existing paths between a pair of servers, i.e., source $s$ and destination $d$, are known. Here, an end-to-end delay $\tau_l$ from $\vv \tau$ and the corresponding path occurrence probability $P_l$ from $\vv P$ describe path $p_l$, $l = 0, ..., N-1$, between $s$ and $d$. The delay vector $\vv \tau$ is sorted in the ascending order, i.e., $\tau_{l+1}\ge \tau_{l}$, for any path $p_{l+1}$ and $p_l$. Generally, each packet sent over path $p_l$ between $s$ and $d$, traverses $\mathfrak y_{l}$ intermediate (forwarding) nodes $v_{q}\in p_l$, $1\leq q\leq \mathfrak y_{l}$. Thus, an ideal mean FCT $t'$ over all existing paths is defined as
\begin{equation}\label{idealFCT}
t^{ideal}=  \sum_{l=0}^{N-1} (\mathfrak y_{l}+1)\frac{L_{flow}}{c}\cdot\frac{P^u_l}{P_U}
\end{equation}
, where $L_{flow}$ and $c$ are flow length in bits and link bit rate, respectively, while probability $P^u_l$ that path $p_l$ is utilized, depending on the scheme. To assess how frequently a path $p_l$  is utilized in comparison to other paths, the path utilization probability $P^u_l$ is normalized by probability $P_U$ that at least one out of $N$ paths is utilized:
\begin{equation}\label{idealFCT}
P_U=  \sum_{l=0}^{N-1} P^u_l
\end{equation}

In case of ECMP, one random path is selected for the transmission of a flow. Thus, the probability, that an arbitrary path $p_l$ is utilized, is defined as:

\begin{equation}\label{PrPath}
P^u_l =\frac{1}{N}
\end{equation}
Thus, the ideal mean FCT $\bar{\tau}_{ECMP}$ for ECMP is a function of path utilization probability $P^u_l$, end-to-end delay $\tau_l$ and FCT $t^{ideal}$:
\begin{equation}\label{FCTopt}
\bar{\tau}_{ECMP}=t^{ideal} +\frac{\sum_{l=0}^{N-1} P^u_l\cdot \tau_l}{P_U}
\end{equation}
%, where $t\rq{}$ is flow transmission time, which depends of flow size and transmission rate.
\par In case of RPS, each node $v$ sends packets using $|\mathcal E_{out}(v)|$ equally probable outgoing links. As a result, the packets of a certain flow are randomly spread over all $N$ existing paths towards destination. Thus, all $N$ paths are always utilized, i.e, with the same equal probability $P^u_l=P_l=1$. As a result, the ideal mean FCT $\bar{\tau}_{RPS}$ for RPS is defined as:
\begin{equation}\label{FCTrnd}
\bar{\tau}_{RPS}=t'+ \tau_{N-1}
\end{equation}
, where $\tau_{N-1}$ is the end-to-end delay of the longest utilized path $p_{N-1}$. Without loss of generality, each forwarding node $v$ can forward $y$ incoming packets without queuing delay over set of $|\mathcal E_{out}(v)|\geq y$ outgoing links. That is feasible when the number of input links of node $v$ is less or equal to the number of available output links, i.e. $y\leq|\mathcal E_{in}(v)|\leq|\mathcal E_{out}(v)|$. However, network can contain intermediate node $v$, where the number of available input ports is larger than the number of available output ports, i.e. $|\mathcal E_{in}(v)|\geq |\mathcal E_{out}(v)|$. When a forwarding node $v$ receives $y$ packets, $|\mathcal E_{out}(v)|<y\leq |\mathcal E_{in}(v)| $, $y - |\mathcal E_{out}(v)|\geq 0$ packets cannot be forwarded immediately and thus must be buffered with probability $P_B(v)$.

\par Generally, the source $s$ and an intermediate node $v$ are connected by maximal $N_{sv}$ paths, which can deliver at most $y=|\mathcal E_{in}(v)|$ packets from at most $|\mathcal E_{in}(v)|$ out of $N_{sv}$ paths simultaneously. Thus, there are $a''=\sum_{i=1}^{|\mathcal E_{in}(v)|- |\mathcal E_{out}(v)|}\binom{N_{sv}}{|\mathcal E_{out}(v)| +i }$ paths combinations, which lead to queuing of $y - |\mathcal E_{out}(v)|$ packets in $v$. 
Next, let's set $A_{\alpha}$ be a set of $y$ out of $N_{sv}$ existing paths from collection $\Psi_{sv}$ and a set $B_{\alpha}=\Psi_{sv}\backslash A_{\alpha}=\{p_l | (p_l \in\Psi_{sv}) \land (p_l \notin A_{\alpha}) \}$ a set of paths, which are not delivering packets at the considered time point. The probability $P''(\alpha, y , N_{sv} )$, $y =|A_{\alpha}| $, $1\leq y \leq |\mathcal E_{in}(v)| $, that paths from $A_{\alpha}$ and not from $B_{\alpha}$ deliver $y$ packets simultaneously is defined as follow
\begin{equation}\label{combProb1}
P''(\alpha, y, N)=\prod_{i=1}^{y} P_{l,i}\prod_{j=1}^{N-y}(1-P_{l,j})
\end{equation}
%\begin{equation}\label{combProb}
%P_B(\alpha)=\prod_{i=1}^{| \mathcal E_{in}(v)| } \prod_{j=1}^{\Phi_{sv}-| \mathcal E_{in}(v)| }P_{l,i}(\alpha)\cdot  (1-P_{l,j}(\alpha))
%\end{equation}
, where $P_{l,i}$ and $P_{l,j}$ are path occurrence probabilities from $\vv P$ collected in set $A_{\alpha}$ and $B_{\alpha}$ and indexed by $i$ and $j$, respectively. The resulting queuing probability in an arbitrary forwarding node $v$ can be derived by considering all $a''$ blocking path combinations as
\begin{equation}\label{blocking}
P_B(v)=\sum_{\alpha=1}^{a''} P''(\alpha, | A_{\alpha}| , N_{sv} )
\end{equation}

As a result, when more than one packet claims the same link at the same time in an intermediate node $v_{q}\in p_l$, $1\leq q\leq \mathfrak y_{l}$, on path $p_l$, and, thus, need to be buffered, each packet can be delayed with probability $P_B (v_{q})$ for $\varrho_q$ \emph{time units}. Thus, the resulting mean FCT $\bar{\tau}_Q$ over all $N$ existing paths is
\begin{equation}\label{pathDelay}
\bar{\tau}_Q= \bar{\tau} + \sum_{l=0}^{N-1}\sum_{q=1, \atop v_{q}\in p_l }^{\mathfrak y_l}P_B(v_{q}) \cdot \varrho_q 
\end{equation}
, where $\bar{\tau}:=\bar{\tau}_{ECMP}$ and $\bar{\tau}:=\bar{\tau}_{RPS}$ for ECMP and RPS, respectively. We assume that queuing delay $\varrho_q$ in each node and the probability $P_{loss}$, that a sent packet is dropped due to buffer overflow in intermediate node and, thus, needs to be retransmitted, can be measured. When considered data flow consists of $H$ packets, the probability $P_r$ that at least one packet of this flow is dropped and retransmitted is defined as
\begin{equation}\label{Pr}
P_r= 1-(1-P_{loss})^H
\end{equation}
The need for retransmission increases resulting FCT $\bar{\tau}_{\sum}$ as 
\begin{equation}\label{FCTcompl}
\bar{\tau}_{\sum}= (1-P_r)\bar{\tau}_Q+P_r\bar{\tau}_Q\cdot\gamma
\end{equation}
, where $\gamma>1$ is a proportion how much FCT is increased as compared to $\bar{\tau}_Q$. In the next section, we compare the expected values of $\bar{\tau}_{\sum}$ (FCT) for the three schemes in order to verify the correctness of our solution.

\begin{table}[!t]
	\renewcommand{\arraystretch}{1.3}
	\caption{Simulation parameters}
	\label{parameters}
	\vspace{-0.1cm}
	\centering
	\begin{tabular}{c|c}
		\hline
		\textbf{Parameter} & \textbf{Value}\\
		\hline
		%General service time  & 3 $\mu s$\\
		Service time (core/aggre. switches) & 12 $\mu s$ - 3 $\mu s$\\
		%Switch queue type & DropTail\\
		Sending rate for each interface & 1 Gbps\\
		Queue length & 1.5 MBytes \\
		Packet size & 1500 Bytes\\
		Short flow sizes & [1.5 , 15] KBytes\\
		Long flow size & [15, 1500] KBytes\\
		\hline
	\end{tabular}
\vspace{-0.5cm}
\end{table}

\section{Performance evaluation} \label{performance}

In this section, using a simulator implemented in JAVA, we evaluate and compare DiffFlow with traditional ECMP and RPS schemes. Since our solution is open to any transport protocol and for simplicity reasons in the analysis and simulations, we obviate the use of TCP and we only retransmit a lost packet after RTT time. Therefore, we focus on packet routing depending whether, given a certain flow, all the packets follow the same randomly chosen path (i.e. ECMP) or every packet is randomly sprayed over all possible paths (i.e. RPS). On the other hand, our simulator use two independent queues for each kind of flow, one for short flows and another for long flows. Using TCP or any other transport protocol will vary the results in relation with our simulations, but still our solution maintains valid, since it is working in layer 2-3.

In the analyzed topology (see Fig. \ref{hybrid}), each server has 8, 2 or 1 possible paths depending if it wants to connect to a server at a different pod, at a different TOR but in the same pod or directly connected to the same TOR, respectively. The determination of the flow lengths follows two independent Poisson processes with mean values 10 and 1000 KBytes for short and long flows respectively. However, the size of short flows is constrained between 1.5 KBytes and 15 KBytes, while the size of long flows goes from 16.5 KBytes to 1500 KBytes. With a fixed packet size of 1500 Bytes, short flows can have a length between 1 and 10 packets, while long flows have a length between 11 and 1000 packets, with mean values of 6 and 666 packets, respectively. On the other hand, the 90\% of the incoming flows are short flows, while the remaining 10\% are long flows. The generation of incoming flow requests follows a Poisson Process with load varying from 0.1 to 0.8 and the service time of switches is fixed to a deterministic value. The rest of parameters can be found in Table \ref{parameters}.

Following, we compare the simulations results for FCT with the values obtained from analysis in Section \ref{analysis}, for each kind of flow separately. Generally, the measured values from simulations follow the analysis, except a deviation occurred due to the chosen value of $\gamma$, $\gamma=5$ for short flows and $\gamma=1.2$ for long flows, which is related with how much the retransmission time affect the FCT. However, with the same $\gamma$ value for the three methods, we can see that the behavior is similar and the simulation results are close to analysis.  

\begin{figure*}[!t]
	\centering
	\subfloat[Mean FCT]{\includegraphics[width=0.9\columnwidth]{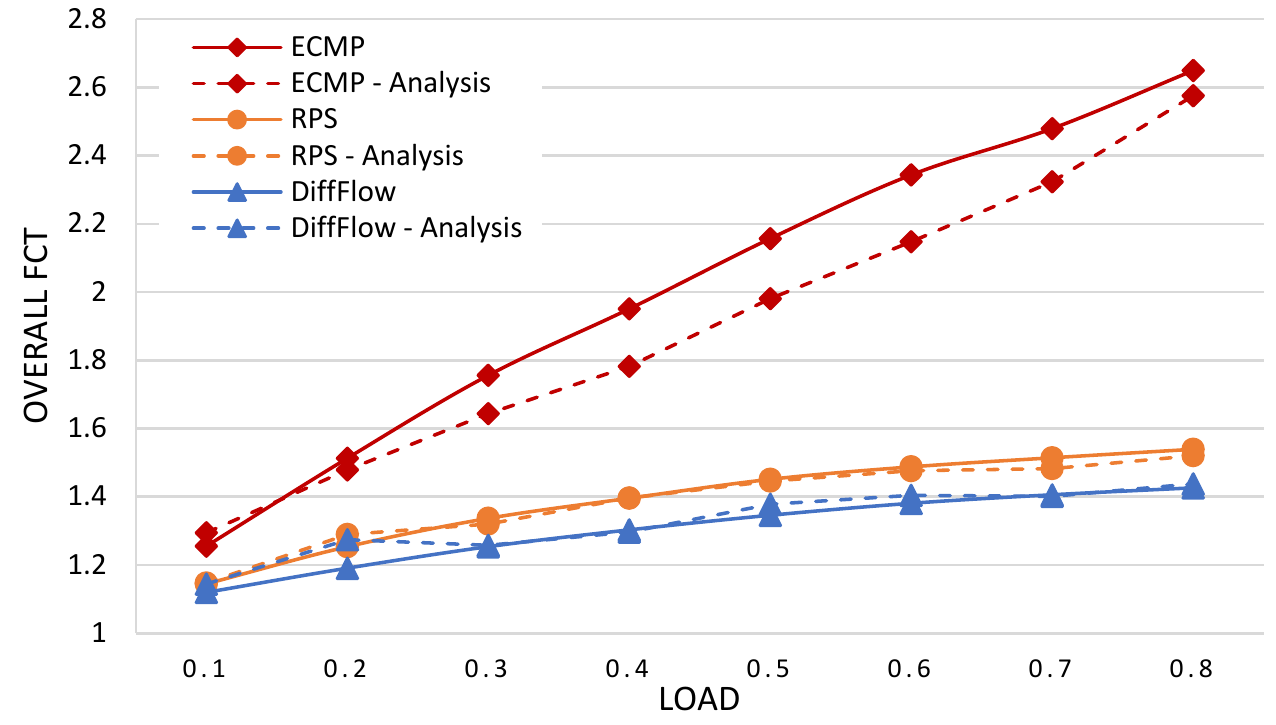}
		\label{overall}}
	\hfil
	\subfloat[Throughput]{\includegraphics[width=0.9\columnwidth]{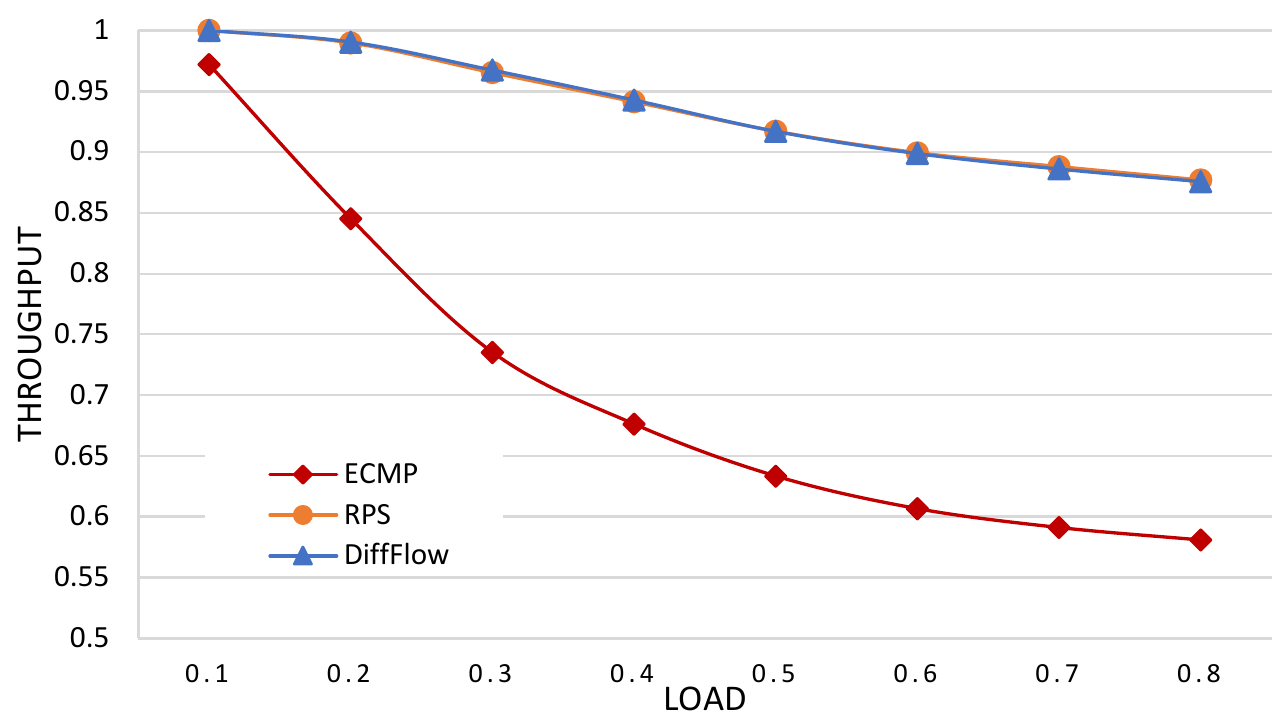}
		\label{throughput}}
			\vspace{-0.35cm}
	\caption{Overall performance}
	\label{Overall_performance}
	\vspace{-0.4cm}
\end{figure*}

\begin{figure*}[!t]
	\centering
	\subfloat[Short flows]{\includegraphics[width=0.9\columnwidth]{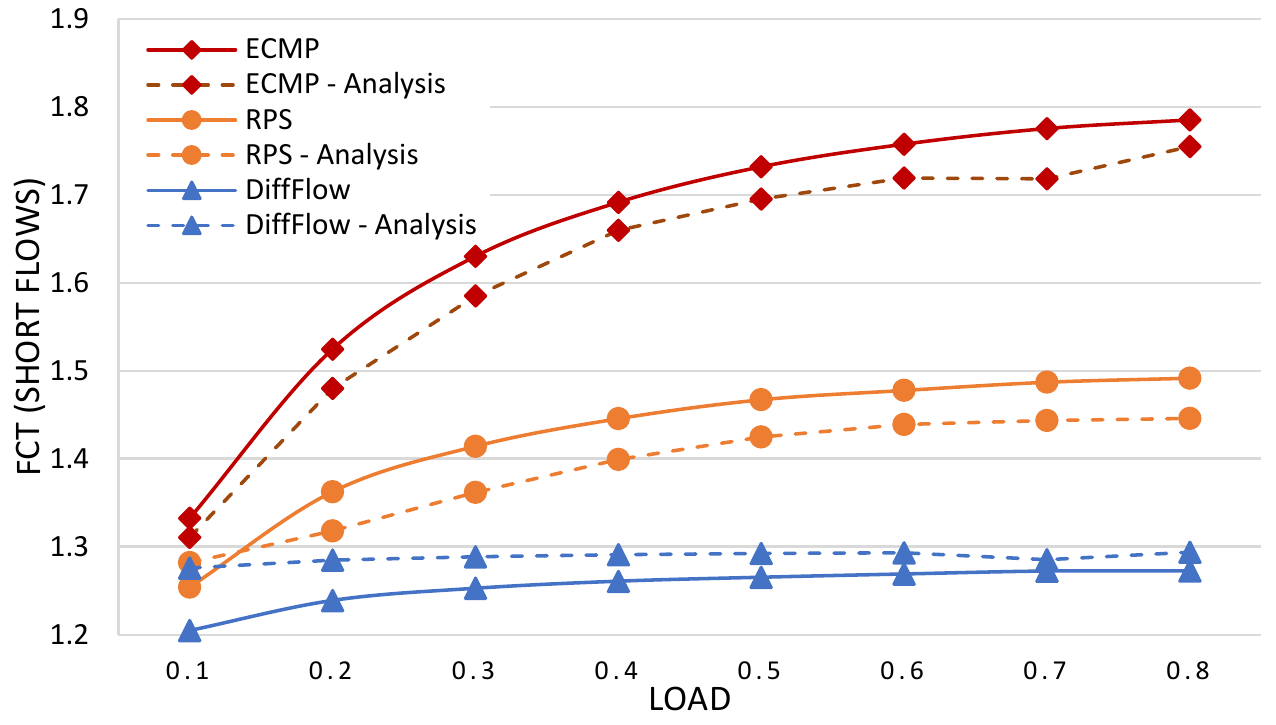}
		\label{short_flows_fct}}
	\hfil
	\subfloat[Long flows]{\includegraphics[width=0.9\columnwidth]{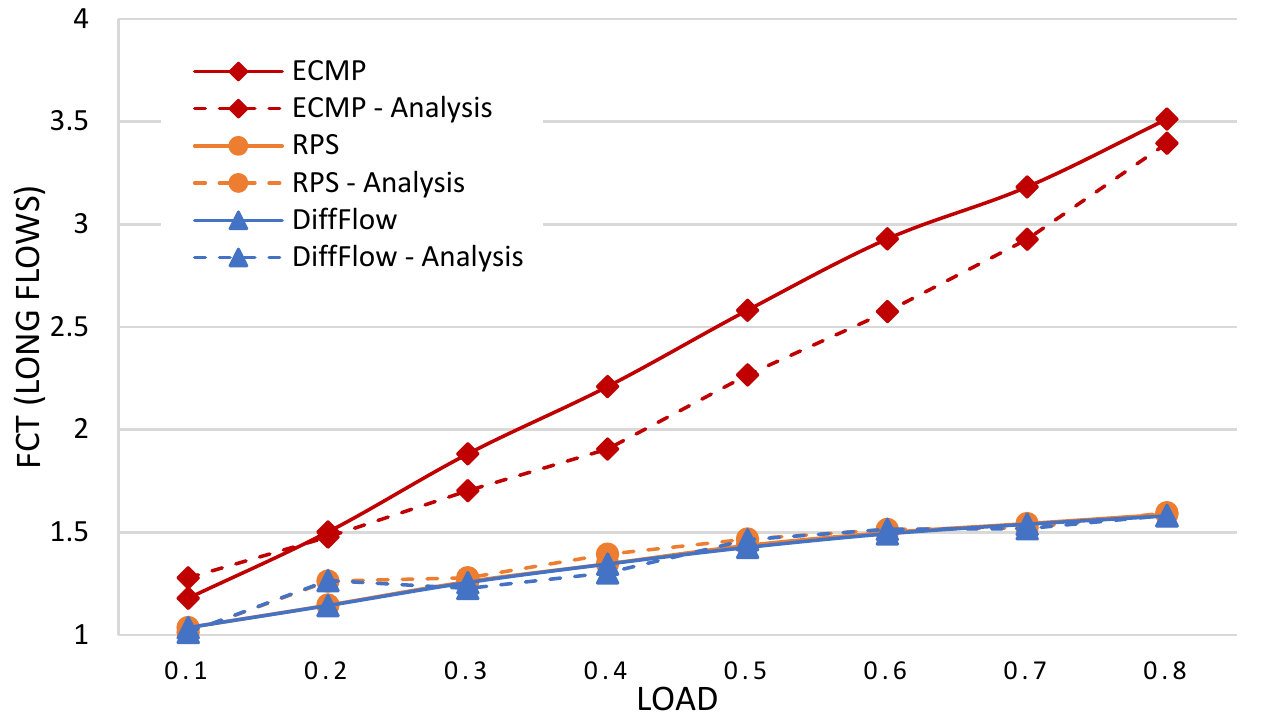}
		\label{long_flows_fct}}
		\vspace{-0.35cm}
	\caption{Mean Flow Completion Time}
	\label{Flow_Completion_Time}
	\vspace{-0.4cm}
\end{figure*}

\subsection{Overall Performance}

As we described in Section \ref{performance_metrics}, the goal of our performance is to minimize the FCT for short flows, maximizing the throughput. The presented figures show the normalized results.

In Fig \ref{overall}, we show how our solution improves to RPS and ECMP in terms of FCT for all flows. While, ECMP is affected when the network load is high due to the creation of hot-spots, RPS and DiffFlow maintain a reasonable FCT, corroborating our expectation, with small variations. Although, RPS is a better solution than ECMP especially for a high traffic load (0.8), the proposed DiffFlow improves RPS by around 7\%, whereby there is no out-of-order problem for short flows.

In terms of throughput, the Fig. \ref{throughput} shows the averaged results for the three schemes for all flows. Since in this case, the percentage of traffic associated to short flows is negligible in comparison to long flows, we can not appreciate improvements of throughput by DiffFlow in comparison with RPS. Both methods result in decrease of throughput from 99.9\% to 87\%, when the traffic load increases from 0.1 to 0.8 Erlang, respectively. In contrast, throughput of ECMP decreases up to 58\% for load 0.8. %However, we can assure that the proposed solution is at least as good as RPS, but in this case, with the same out-of-order issue than RPS. 

\subsection{Performance of Short Flows} \label{shortFlows}
 
Fig. \ref{short_flows_fct} shows the comparison of methods regarding FCT for short flows. Here, RPS improves ECMP up to 20\%, while proposed DiffFlow improves RPS by around 15\% when the network load is higher than 0.3. That is due to the fact, that, in case of RPS, packets of all flows are randomly spread over all existing paths, therefore, packets of short flows are perceiving different path latencies depending on the chosen path, and this directly affects the FCT. However, short flows in DiffFlow are using ECMP, avoiding this problem, and long flows are not creating bottlenecks in the network, as compared to ECMP for all flows. Therefore, we demonstrate that the use of RPS for short flows is not as efficient as DiffFlow, and even the out-of-order problem is presented in the RPS case.

\subsection{Performance of Long Flows}

In Fig. \ref{long_flows_fct}, the FCT of long flows for all discussed transmission methods is presented. However, RPS has a lot of benefits in comparison with traditional ECMP due to the efficient load balancing method. Here, we can see that the FCT for ECMP increases quickly from 1.2 up to 3.5 with increasing network load, while RPS and DiffFlow increase slightly from 1 to 1.5. The out-of-order problem, present in both RPS and DiffFlow, is not an important aspect for long flows, since they do not have temporal requirements. Moreover, newer versions of TCP, with DSACK and timestamps options, are more robust to packet reordering, minimizing this problem.

\section{Conclusion}

We presented a new multipath routing scheme for DCNs that differentiates between the routing method used for short and for long flows. Making use of SDN and packet sampling technologies, our approach can detect and forward long flows using random packet spraying, while short flows are forwarded using ECMP. As we demonstrate, DiffFlow improves the FCT for short flows and throughput for long flows in comparison with traditional ECMP. However, the need to use a centralized controller for the advertisement of long flows, makes that our solution cannot be applied in current DCNs, but we encourage to follow researching on the  future SDN architecture.


%\section*{Acknowledgment}




%\bibliographystyle{IEEEtran}

\bibliographystyle{IEEEtran}
\bibliography{mylib}

\end{document}